\documentclass[sn-mathphys,Numbered]{sn-jnl}

\usepackage{graphicx}%
\usepackage{multirow}%
\usepackage{amsmath,amssymb,amsfonts}%
\usepackage{amsthm}%
\usepackage{mathrsfs}%
\usepackage[title]{appendix}%
\usepackage{xcolor}%
\usepackage{textcomp}%
\usepackage{manyfoot}%
\usepackage{booktabs}%
\usepackage{algorithm}%
\usepackage{algorithmicx}%
\usepackage{algpseudocode}%
\usepackage{listings}%
\usepackage[colorinlistoftodos,textwidth=50,textsize=tiny]{todonotes}

\usepackage{soul}
\usepackage{ulem}

\begin{document}

\newcommand\green[1]{\textcolor{green}{#1}} 
\setstcolor{green} 
\newcommand\purple[1]{\textcolor{purple}{#1}} 
\setstcolor{purple} 
\newcommand{\CA}[1]{{\color{teal}#1}} 
\setstcolor{teal} 

\title[Article Title]{Phase Retrieval of Highly Strained Bragg Coherent Diffraction Patterns using Supervised Convolutional Neural Network}


\author[1,2,*]{\fnm{Matteo} \sur{Masto}}
\equalcont{These authors contributed equally to this work.}

\author[1,2]{\fnm{Vincent} \sur{Favre-Nicolin}}
\author[1]{\fnm{Steven} \sur{Leake}}
\author[1]{\fnm{Tobias} \sur{Schülli}}
\author[3]{\fnm{Marie-Ingrid} \sur{Richard}}
\author[1]{\fnm{Clement} \sur{Atlan}}
\author[4,*]{\fnm{Ewen} \sur{Bellec}} \email{matteo.masto@esrf.fr, ewen.bellec@esrf.fr}
\equalcont{These authors contributed equally to this work.}

\affil[1]{\orgdiv{ESRF}, \orgname{The European Synchrotron}, \orgaddress{\street{Av. de Martyrs, 71}, \city{Grenoble}, \country{France}}}

\affil[2]{\orgname{Univ. Grenoble - Alpes}, \orgaddress{\city{Grenoble}, \country{France}}}

\affil[3]{\orgname{Univ. Grenoble Alpes, CEA Grenoble, IRIG, MEM, NRX}, \orgaddress{\street{17 rue des Martyrs}, \city{Grenoble}, \country{France}}}

\affil[4]{\orgname{Univ. Grenoble Alpes, CNRS, Grenoble INP, Institut Néel}, \orgaddress{\street{25 rue des Martyrs}, \city{Grenoble}, \country{France}}}
\maketitle

In Bragg Coherent Diffraction Imaging (BCDI), Phase Retrieval of \textit{highly} strained crystals is often challenging with standard iterative algorithms. This computational obstacle limits the potential of the technique as it precludes the reconstruction of physically interesting highly-strained particles. Here, we propose a novel approach to this problem using a supervised Convolutional Neural Network (CNN) trained on 3D simulated diffraction data to predict the corresponding \textit{reciprocal space phase}. This method allows to fully exploit the potential of the CNN by mapping functions within the same space and leveraging structural similarities between input and output. The final object is obtained by the inverse Fourier transform of the retrieved complex diffracted amplitude and is then further refined with iterative algorithms. We demonstrate that our model outperforms standard algorithms on highly strained simulated data not included in the training set, as well as on experimental data.




\section{Introduction}\label{sec1}

\begin{figure}[h]
    \centering
    \includegraphics[width=\textwidth]{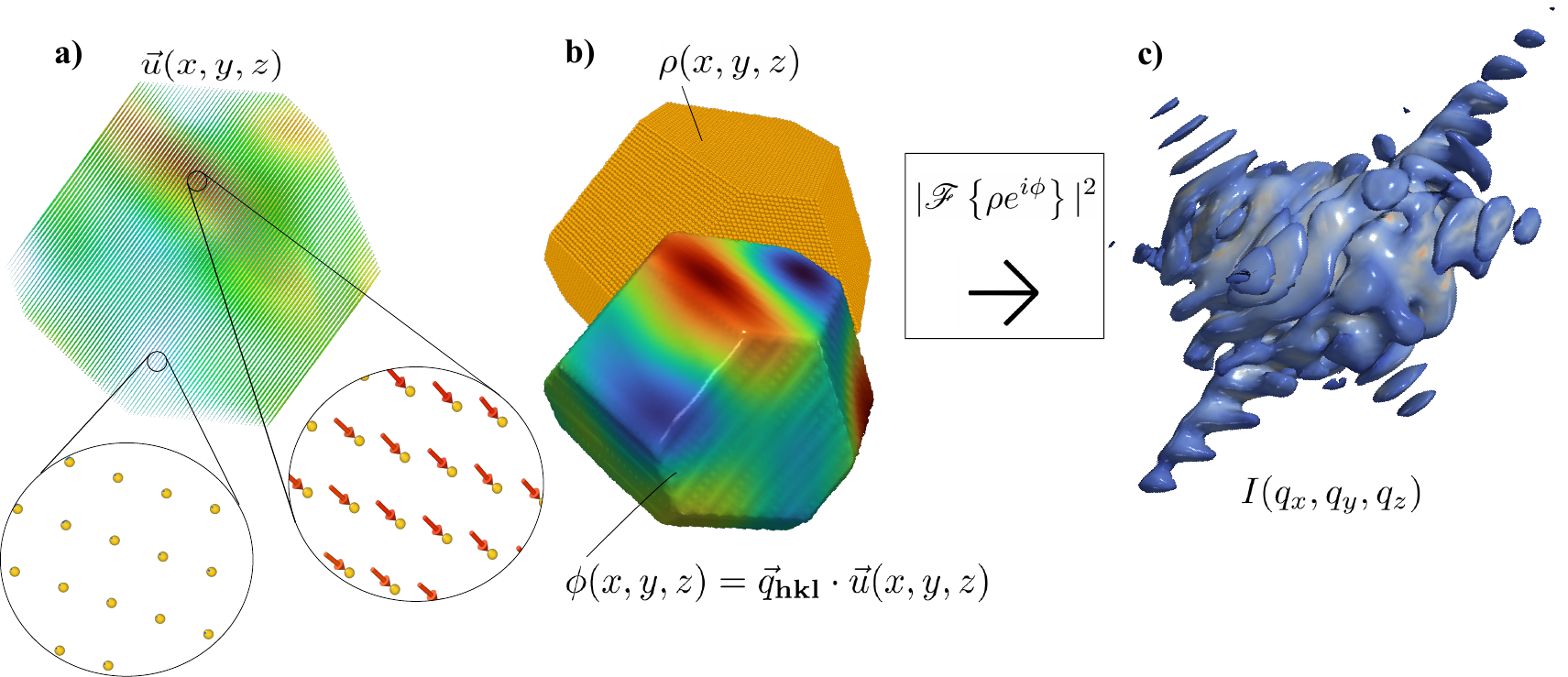} 
    \caption{\textbf{a)} Simulated displacement field $\vec{u}(x,y,z)$ in a gold particle with Winterbottom shape and zoomed regions with small (\textit{left}) and large (\textit{right}) atomic displacements. \textbf{b)} At the exit plane of the particle, the complex diffracted wave is composed of a modulus and a phase. While the former - $\rho(x,y,z)$ - represents the electron density, the latter - $\phi(x,y,z)$ - is given by the projection along the scattering vector $\vec{q}_{\textbf{hkl}}$ of the displacement field. \textbf{c)} In the far-field regime the diffracted wave is Fourier transformed and only its squared modulus is collected while the reciprocal space phase is lost.}
    \label{fig:displacement_phase} 
\end{figure}

Bragg Coherent Diffraction Imaging (BCDI) is an X-ray microscopy technique performed at synchrotron and XFEL facilities to image lattice deformations within nanocrystalline structures with spatial resolution down to a few nanometers \cite{miao_approach_2001, Robinson2009, Richard:te5091}. The primary reason for using BCDI is indeed its sensitivity to strain and crystallographic defects \cite{Carnis2021, PhysRevMaterials.4.013801}. In fact, any internal lattice deformation of the sample significantly affects the measured diffraction pattern. In a standard BCDI experiment, the sample is fully illuminated by a coherent X-ray beam and the scattered photons are collected by a pixelated detector placed in the far-field regime, aligned with a Bragg reflection. The acquired intensity represents the squared modulus of the Fourier transform of the probed crystal \cite{Godard2021}.
The reconstruction of the 3D electron density and strain map of the particle from the diffraction pattern is entrusted to computer algorithms that perform the so called Phase Retrieval (PR). At convergence, the algorithms find a complex object the modulus of which is interpreted as the electron density while the phase is the projection of the atomic displacement (relative to the average lattice of the crystal) along the scattering vector  (Fig.\ref{fig:displacement_phase}). Because strain manifests itself through this phase variation, one can often speak of ``strong‑phase'' while referring to ``high-strain''.\\
The technique's non-invasive nature makes it suitable for \textit{in situ} and \textit{operando} measurements in different sample environments and has been successfully employed for electrochemistry studies, gas reactions, high-pressure and high-temperature conditions \cite{Yang2013, IronGold2018, ROCHET2019169, Dupraz2022, Atlan2023, Chatelier:2023}.
\\
PR is the most delicate step of the data processing and it is commonly accomplished by iterative algorithms that project the diffraction pattern alternating between Fourier and direct spaces. After the projection on each space a constraint is applied. In Fourier space the calculated diffraction pattern is replaced by the observed one whereas in direct space, the object's support is updated into a confined and compact volume \cite{ Gerchberg1972APA, Fienup:78, Marchesini2007ARetrieval, Favre-Nicolin2020PyNX:Operators}. While widely corroborated and of standard use in the BCDI community, these algorithms often struggle to find convergence when the strain inside the particle is large, typically resulting in object's phase ranges larger than  2$\pi$.
The distortion of the diffraction pattern caused by the high strain leads to a bad estimate of the particle's unknown support, hindering convergence to the solution. In practice, one needs to try several of randomly initialized reconstructions and combine the best ones to obtain a good solution \cite{Favre-Nicolin2020}. For high-strain cases this process is time consuming as many trials are required to achieve a good reconstructed object. In worst cases multiple PR runs are not sufficient and the reconstruction is not possible. Moreover, it is worth emphasizing that these occurrences are relatively common in BCDI experiments, and the impossibility to attain successful reconstructions because of the high-strain is a major limitation of the technique to explore interesting physical phenomena that involve highly strained particles.

For these reasons, numerous studies have been conducted in order to overcome this problem. Zhao \textit{et al.} exploited a 2D modulator plate between the sample and the detector to modulate the diffraction pattern and introduce an additional constraint during the PR \cite{Zhao, Zhao_unpublished}. Wang \textit{et al.} imposed the same support of the unstrained particle during the particle's phase transformation, but this method is limited to initially unstrained crystals \cite{Wang_2020}. Different computational approaches involved compressive sensing \cite{NewtonStrain} or additional constraints on the object \cite{Minkevich2008}. More recently, Convolutional Neural Networks (CNN) have also been successfully implemented for BCDI phase retrieval. First, Cherukara \textit{et al.} in 2018 \cite{Cherukara2018Real-timeNetworks} successfully inverted 2D simulated BCDI datasets with two CNNs for support and phase prediction, opening the field of deep learning (DL) for end-to-end BCDI phase retrieval. In 2020, Scheinker and Pokharel \cite{Scheinker2020AdaptiveImaging} managed to reconstruct the first 3D experimental BCDI patterns with the help of a CNN for the prediction of the coefficients of the spherical harmonics used to represent the object's support surface. 
Wu \textit{et al.} \cite{Wu:cw5029} opted for an architecture with two parallel decoders for the prediction of the object's modulus and phase, respectively. Their model was able to take as input the 2D central slice of 3D simulated and experimental BCDI patterns and return the corresponding 2D object's projection, also in presence of `strong phases'. The authors also considered the refinement of the DL prediction with iterative algorithms. In 2021 Chan \textit{et al.} extended the encoder/2-decoders architecture to the 3D case, introducing an automatic-differentiation based refinement of the DL predicition. In the same year, the work of Wu and coauthors \cite{ Wu2021Three-dimensionalNetworks} explored the possibility of unsupervised learning, leveraging the speed and robustness against noise of the combined approach. Yao \textit{et al.} published AutoPhaseNN in 2022 \cite{Yao2022AutoPhaseNN:Imaging}, a fully unsupervised physics-informed CNN for BCDI phase retrieval. Yu \textit{et al.} \cite{Yu2024} recently developed a complex-CNN for 2D coherent diffraction patterns which employs complex convolutions, to account for the complex nature of the solutions. Another work leveraging the tensor computed back-propagation is the one of Maddali and coauthors \cite{Maddali2023}, in which automatic differentiation is successfully exploited for the PR of multi-peak BCDI. However, none of these works explicitly mentions results on highly strained 3D experimental data that could not be inverted with conventional PR. \\
This work presents a CNN from a novel perspective as it focuses on the prediction of the reciprocal‑space phase rather than the direct‑space object, with the specific aim of enabling the model to handle high‑strain. In our case, the object is retrieved at a later stage by an inverse Fourier transform of the diffracted complex amplitude composed of the square root of the diffracted intensity and the predicted phase. We believe that this approach is beneficial for two main reasons. Firstly, we leverage the structural similarities between the input diffracted intensities and the corresponding phases, making best use of the feature extraction and transposing capabilities of the adopted UNet-like architecture \cite{3DUnet2016, 3DUnetSeg2023}. In this framework, the use of skip connections, responsible for the combination of local information with context information of the encoder and decoder respectively \cite{Xu2024}, is also optimized when input and output images belong to the same space, real or reciprocal. Lastly, we reduce the complexity of the problem by predicting a single array (reciprocal space phase) instead of two different and coupled ones (object modulus and phase). 

\section{Results}\label{sec2}
\subsection{Dataset preparation}
The model training is fully supervised and this choice, a minor, necessary compromise, can be supported by the limited variety of single-crystal sample shapes typical of the BCDI technique. However this implies the simulation of thousands of intensity-phase pairs that would be used as an input-ground truth reference. We started by creating 3D particles of different shapes - Wulff, Winterbottom, and random ones generated by making planar cuts of various sizes and orientations from a cube - following the procedure used by Lim \textit{et al.} in Ref. \cite{Lim2021ADiffraction}. From each particle, several diffraction patterns were calculated changing the oversampling ratio (defined per spatial dimension as the quotient of computational‐array size to crystal size, which must exceed 2 in accordance with the Nyquist criterion to guarantee invertibility), the rotation of the reciprocal space coordinates and the strain distribution. The particles' strain distribution was simulated applying an artificial complex phase to the each crystal shape. This was either obtained with two Gaussian functions, or two cosine functions or with random Gaussian distribution \cite{Wu:cw5029}. For each case, amplitude, variance, frequency, and correlation length were randomly chosen to ensure a phase variation within the particle ranging between $2\pi$ and $5\pi$. The diffraction patterns were computed using the PyNX scattering package \cite{Favre_scattering} on a cubic, 64 pixel-side, $q_x, q_y, q_z$ grid. As last step, noise from Poisson statistic was applied to the signal in order to simulate the experimental conditions (see also \cite{Masto:yr5131} Supp. Material).
A total number of 95,000 three-dimensional 64 pixel-size diffraction patterns were used for training, 4,000 for validation and 3,000 for testing.

\begin{figure}[h]
    \centering
    \includegraphics[width=\textwidth]{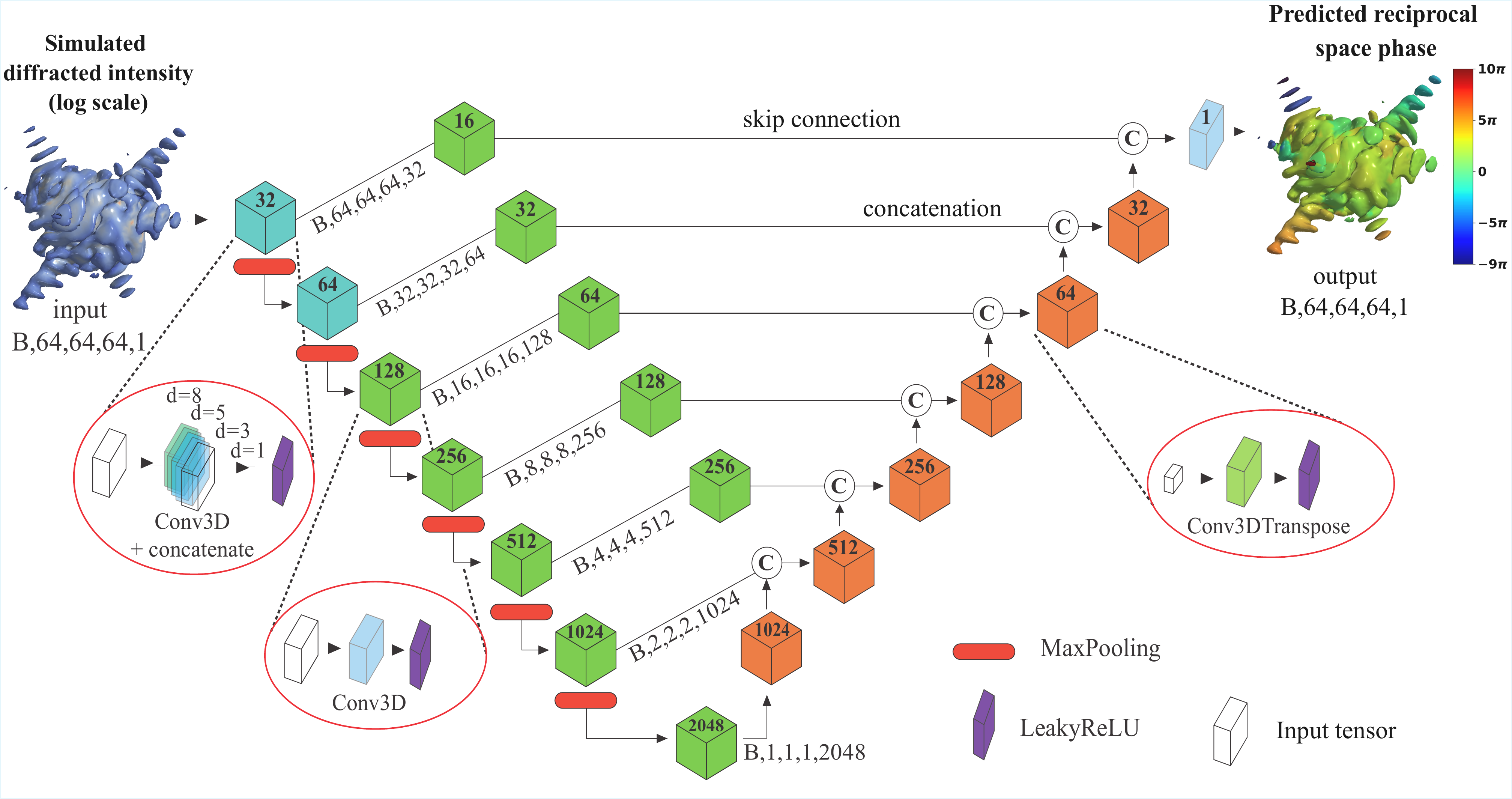} 
    \caption{Schematic of the network architecture. From the left, batches (B) of 3D log-scaled BCDI intensities are sent through the encoder and reduced to line vectors in the bottleneck. The decoder, concatenating the outputs of the skip blocks with each current state, reshapes the output to the original size. The reciprocal space phase, predicted without last activation layer is automatically delivered in the unwrapped form.}
    \label{fig:architecture} 
\end{figure}

\subsection{Network architecture and training}
The network architecture (Fig. \ref{fig:architecture}) is inspired by the 3D UNet model for image segmentation \cite{3DUnet}. The encoder takes as inputs the diffracted intensities in logarithmic scale normalized between 0 and 1 and process them with a series of convolutional and max pooling layers. The reason behind the transformation into logarithmic scale is twofold. First, it compresses the range of intensity values, reducing the disparity between the central peak and higher-order fringes, which helps the model better capture features associated with higher $Q$ values. Second, the spatial structure of the reciprocal space phase aligns more closely with the logarithm of the intensity, promoting a more efficient flow of information from the encoder to the decoder (blue/green and orange blocks respectively in Fig. \ref{fig:architecture}). Copies of the tensors after each convolutional block are sent through other convolutional layers outside the encoder and used as skip connections. The decoder starts from the smallest feature map and by means of transposed convolutions enlarges the tensor shape up to the original one, for a total of around 143 millions of trainable parameters. No activation function was applied to the last convolutional layer allowing the model to produce unbounded outputs, thereby accommodate for all phase symmetries (see next Section).   The first two blocks, already successfully used in our previous work \cite{Masto:yr5131}, combine convolutions computed for different dilation rates, improving the receptive field and the ability to extract longer-range spatial correlations \cite{Chen2017RethinkingSegmentation}. The network is written using the Tensorflow Python library \cite{abadi2016tensorflow} and trained for 60 epochs ($\sim$ 30 hours), loading batches of 16 images at a time, using the ADAM optimizer \cite{kingma2017adam} with a learning rate of $10^{-4}$ on two NVIDIA Tesla V100-SXM2-32GB GPUs.

\subsection{Loss function}
The prediction of the reciprocal space phase $\varphi(\vec{q})$ (where $\vec{q}$ is the coordinates vector in reciprocal space), poses some challenges for what concerns the choice of the training loss function. In fact, common metrics like the Mean Squared Error (MSE), Mean Absolute Error (MAE) or Structural Similarity (SSIM) cannot account for the phase symmetries, that in our case are:
\begin{itemize}
    \item  \textbf{ramp}: $\varphi(\vec{q})$ and $\varphi(\vec{q}) + \vec{q} \cdot \vec{x}_0 $ are equivalently good solutions. The term $\vec{q} \cdot \vec{x}_0  $ appears when the object is displaced from the center of the reference frame by a quantity $ \vec{x}_0 $ in direct space, and this information is not encoded in the diffraction pattern.
    \item  \textbf{offset}: $\varphi(\vec{q})$ and any other $\varphi(\vec{q}) + c$, where $ c \in$ $\mathbb{R}$ is a constant value across the whole phase, are equivalently good solutions.
    \item  \textbf{phase wrap}: given a position $q_0$ in $q$-space, the phase $\varphi(q_0)$ and any other $\varphi(q_0) + 2k\pi$ (with $k$ $\in$ $\mathbb{Z}$) are equivalent. 
    \item  \textbf{sign}: $\varphi(\vec{q})$ and $-\varphi(\vec{q})$ are equivalently good solutions. The sign affects the orientation of the particle in real space, discriminating between the two possible twin solutions. This information is not encoded in the diffraction pattern.
    
\end{itemize}

While the first symmetry is broken by using a training dataset created from centered particles, thus biasing the model for a $x_0$ = 0, in order to consider the other three symmetries we designed the following tailored loss function that we called Weighted Coherent Average (WCA):

\begin{equation}
    L_\pm = 1 - \left|\frac{1}{N}\sum_{k=1}^{N} I_{\text{input},k}\exp\left(i(\pm\varphi_{\text{GT},k} - \varphi_{\text{pred},k})\right)\right|
\label{eq:1}
\end{equation}

where $N$ is the total number of voxels in each phase array $(64 \times 64 \times 64)$ and $k$ is the voxel index, $I_{\text{input},k}$ is the log-scaled BCDI intensity, $\varphi_{\text{GT}}$ is the ground truth reciprocal phase and $\varphi_{\text{pred}}$ the model prediction. Finally, the phase sign is chosen as : 

\begin{equation}
    L_{\text{WCA}} = \min\left(L_+, L_-\right)
\label{eq:2}
\end{equation}

Equation \ref{eq:1} induces all the voxels ($k$) of the predicted phase ($\varphi^{k}_{\text{pred}}$) to have the same constant offset across the tensor. In fact, maximizing the modulus of the complex average vector of the  complex phase-difference vectors means aligning them to sum \textit{coherently}, thus having the same phase-difference throughout all the voxels. In addition, expressing the phase difference as the angle of a complex exponential naturally removes the wrap problem. On the other hand, weighting the complex phase-difference vectors with the intensity value of the corresponding voxels gives a preferential direction to the gradient descent speeding up the loss minimization (see Supp. S1). Eq \ref{eq:1} includes the $\pm$ symbol, as it should be used for both the positive and negative reciprocal space phases. In Eq. \ref{eq:2}, the smallest of the two values is chosen for the back-propagation. In fact, both are good solutions and since the information about the sign cannot be predicted from the intensity pattern, it needs to be left as degree of freedom to the model. With regard to this, it is worth noticing that any phase $\phi$ composed of $\varphi(\vec{q})$ in some regions of the $q$-space and $-\varphi(\vec{q})$ elsewhere, is not a stable minimum point for Eq.\ref{eq:1} and is thus discarded by Eq. \ref{eq:2}. Solutions like $\phi(\vec{q})$ are at the origin of the twin-image problem in the corresponding reconstructed objects \cite{Guizar-Sicairos:12}, a recurrent issue for conventional iterative algorithms that can affect PR deep learning models as well if not properly resolved \cite{Sun_symmetry2024} .

\subsection{Results in reciprocal space - phase prediction }

\begin{figure}[h!]
    \centering
    \includegraphics[width=\textwidth]{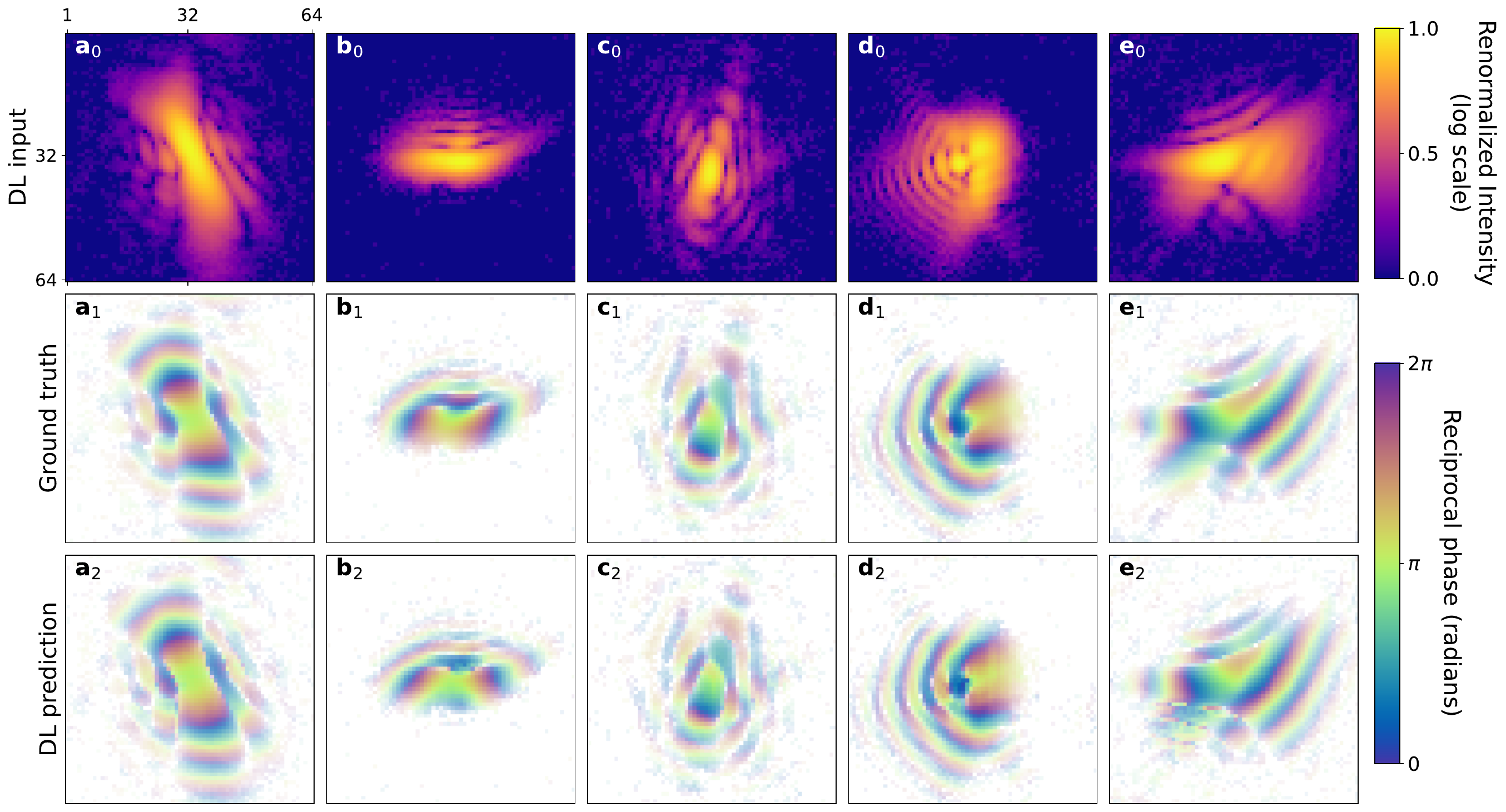} 
    \caption{\textbf{\(\boldsymbol{a_0}\) - \(\boldsymbol{e_0}\))} Central slices of simulated input intensities from the test dataset. \textbf{\(\boldsymbol{a_1}\) - \(\boldsymbol{e_1}\))} Corresponding ground truth reciprocal space phase. \textbf{\(\boldsymbol{a_2}\) - \(\boldsymbol{e_2}\))} Corresponding slices taken from the Deep Learning (DL) model prediction of the reciprocal space phases} 
    \label{fig:sim_phases} 
\end{figure}

Once trained, the model was tested on BCDI noisy patterns of the test dataset. Figure \ref{fig:sim_phases} shows slices of 5 simulated strained BCDI patterns with corresponding ground truth and predicted phases. The predicted phase has been wrapped between 0 and $2\pi$ to have better visual comparison with the ground truths. The model correctly predicts the phase oscillations between fringes and is able to guess the phase where the high strain field has deformed the pattern. 

\subsection{Reconstructions in real-space} 

\subsubsection{Simulated real-space result}
\begin{figure}[h!]
    \centering
    \includegraphics[width=\textwidth]{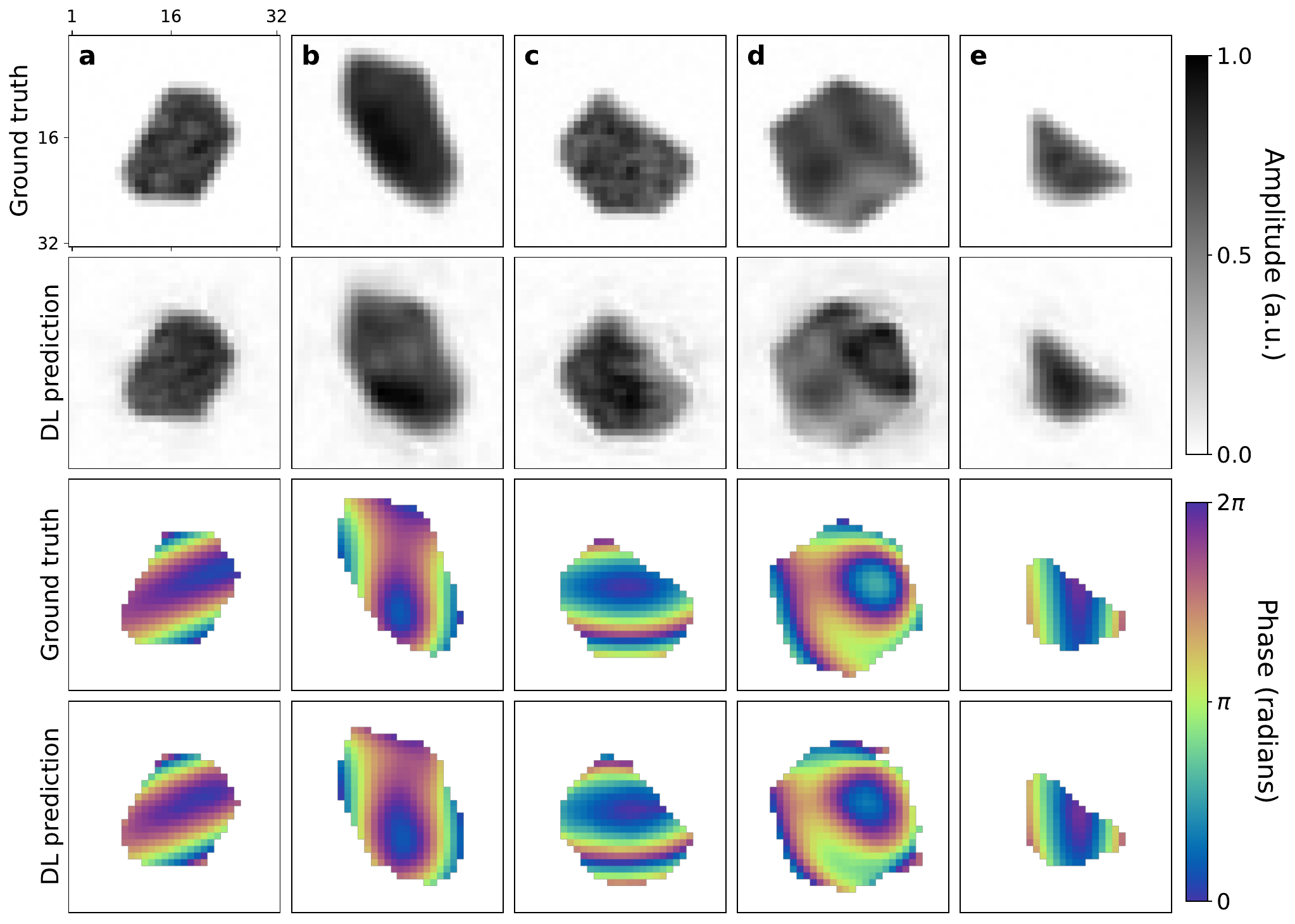} 
    \caption{\textbf{Reconstructed objects from the diffraction patterns in Fig. \ref{fig:sim_phases} }\textbf{a-e)} Central slices of both the modulus and phase for the ground truths and the DL reconstructions.}
    \label{fig:sim_objs} 
\end{figure}

By computing the inverse Fourier transform of the complex diffracted amplitude composed of the modulus (square root of the diffracted intensity) and the phase (predicted reciprocal space phase) one can retrieve the objects in real space. Figure \ref{fig:sim_objs} depicts the ground truth and predicted objects obtained from the simulated diffraction patterns displayed in Fig. \ref{fig:sim_phases}.

Barring some noise and inhomogeneities in the objects' moduli, the reconstructions from the predicted reciprocal space phase achieve good accuracy on simulated data, for different particle shapes and strain distributions. Notably, in BCDI reconstructions, a modulus sufficient for resolving particle's support is adequate whereas the phase within the support, being the carrier of strain information, is fundamentally more significant.

\subsubsection{Results on experimental data}
\begin{figure}[h]
    \centering
    \includegraphics[width=\textwidth]{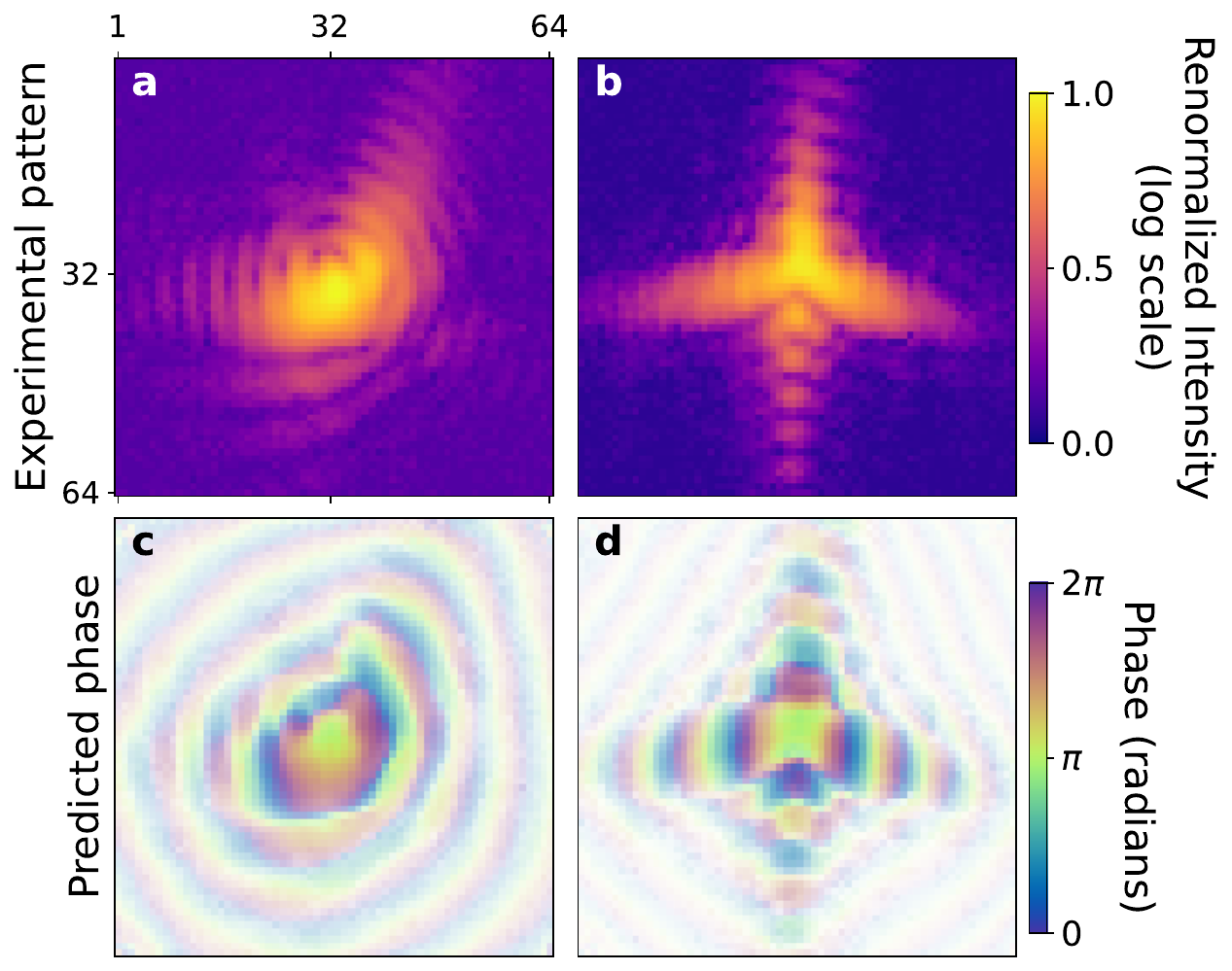} 
    \caption{\textbf{Experimental BCDI patterns. }\textbf{a - b)} Central slices of two experimental BCDI patterns. \textbf{c-d)} Corresponding slices of the DL model phase predictions.}
    \label{fig:exp_data} 
\end{figure}

 Here we discuss the use of our DL model on experimental BCDI data of highly strained particles. We present two relevant examples of BCDI patterns collected at the ID01 beamline of the European Synchrotron Radiation Facility - Extremely Brilliant Source (ESRF-EBS). Figure \ref{fig:exp_data} shows one central slice of the experimental BCDI patterns and the corresponding retrieved phase obtained using our DL model. Particle 1 (Figs. \ref{fig:exp_data}\textbf{a}-\textbf{c}) is a platinum nanoparticle on Yttria-stabilized zirconia (YSZ) substrate \cite{data1} and Particle 2 (Figs. \ref{fig:exp_data}\textbf{b}-\textbf{d}) is a dewetted platinum/palladium bilayer on a sapphire substrate \cite{data2}. These particles have a large strain induced by the particle-substrate interface.
 Since the DL model only accepts cubic datasets with a side length of 64 pixels, the experimental diffraction patterns are firstly cropped around the center of the peak and then resized to $64\times64\times64$ pixels shape for the phase prediction.

\begin{figure}[h!]
    \centering
    \includegraphics[width=\textwidth]{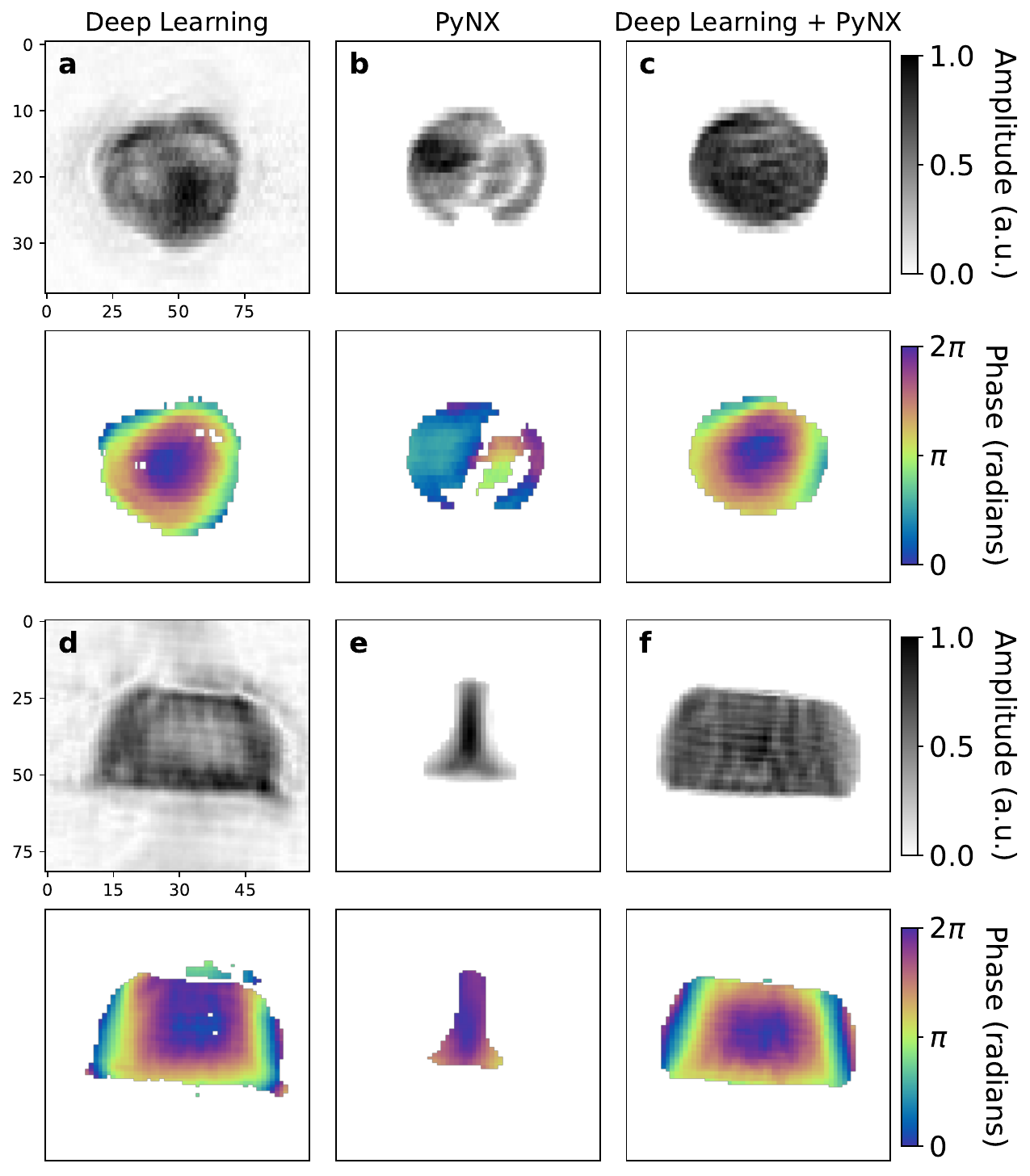}
    \caption{\textbf{Reconstructed objects from the diffraction patterns in Fig. \ref{fig:exp_data} }\textbf{a-d)} Object obtained from the DL prediction of the reciprocal space phase. \textbf{b-e)} Reconstructed object using 400 HIO + 1000 RAAR + 300 ER iterations (best of 60 different reconstructions). \textbf{c-f} Reconstructed object using 400 ER iterations starting from the object obtained from the DL phase prediction.}
    \label{fig:exp_reconstructions} 
\end{figure}

The object obtained from the predicted reciprocal space phase is then re-interpolated back to the original experimental dataset size before being used (Figs. \ref{fig:exp_reconstructions}\textbf{a-d}) as starting guess for iterative PR with PyNX. Providing a first estimate of the object support and phase field enables the standard PR algorithm to reach good convergence within 400 iterations of Error Reduction \cite{Marchesini2007ARetrieval}. It is crucial for this stage that the DL predicted object, despite the noise around the modulus, already approximates well the solution as convergence is found in few error reduction (ER) steps, without excessively deviating, in final shape and phase field, from the DL guess. For this reason, during ER refinement, only the voxels located at the borders of the support are updated, thus avoiding large changes of the object's shape. In fact, it is often the case that in presence of strong object's phase variations, the support shrinking operations \cite{Marchesini2003} during standard PR can erroneously compress the object's support, misleading the algorithm towards bad reconstructions. This is particularly the case of the examples mentioned here as depicted in Fig. \ref{fig:exp_reconstructions}\textbf{e}.

As reference for standard PR, we have chosen the mixed recipe of 400 Hybrid Input Output (HIO) iterations, 1000 of Relaxed Averaged Alternating Reflections (RAAR) and 300 of Error Reduction (ER), varying for each run the threshold parameter for the object support estimated from the auto-correlation of the measured intensity \cite{Crimmins:90}. 

60 independent runs have been launched using \textit{cdiutils} package \cite{atlan2023cdiutils} and PyNX mode decomposition was performed on the three best reconstructions \cite{FreeLLK}. In all cases, the reconstructions were not satisfactory, showing either holes in the object's modulus (Fig. \ref{fig:exp_reconstructions}\textbf{b}) or excessively shrunk support (Fig. \ref{fig:exp_reconstructions}\textbf{e}). This result is particularly important because it marks a milestone for DL‑assisted PR for BCDI: for the first time, it can outperform conventional PR methods, enabling reconstructions of objects with large phase ranges (see Supp. S5) that would otherwise be impracticable. Two other examples of equivalently compared reconstructions of experimental data can be found in Supplementary Material (See Supp. S3)

Besides the higher quality of the reconstructions, one can assess the gain in speed of the DL-PyNX approach compared to classical phasing.
The DL prediction combined with the inverse fast Fourier transform (FFT) is completed within milliseconds, while the PyNX-based ER refinement typically takes 10 to 6 seconds, depending on the dataset size. In contrast, multiple runs - often necessary for inverting challenging datasets, such as those requiring 400 HIO + 1000 RAAR + 300 ER with 20 to 100 independent runs - can extend computation times to several tens of minutes. 
Finally, we have shown that the DL-PR method also benefits low-strain particles, yielding reconstructions of higher quality and greater reproducibility. (Supp. S4).

\section{Conclusion}
In this work we have presented a novel approach for the phase retrieval of highly strained BCDI patterns using a convolutional neural network for the prediction of the reciprocal space phase. We train the CNN in a supervised manner entirely in reciprocal space using our custom Weighted Coherent Average loss on simulated, highly strained BCDI patterns. Nevertheless, our simulations proved to be realistic enough to enable DL phase retrieval of experimental data. In particular, for several high-strain experimental data, the DL reconstructed objects happen to be good initial estimate that can drive conventional phasing to successful reconstructions. To our knowledge, while earlier DL-based PR approaches exclusively accelerated existing routines, our DL model also enables reconstructions that would otherwise be unattainable with conventional PR methods only. 
Additionally, the computational workload is reduced from tens of independent runs, each requiring on the order of $10^2 - 10^3$ iterations, to a single forward inference step followed by approximately  $10^2$ iterations of ER refinement, effecting a two to three orders of magnitude decrease in elapsed real time. Moreover, the DL-PR approach can improve reconstructions of low-strain particles as well. 
Taken together, these advances confer a marked increase in robustness, efficiency, and versatility. 

We therefore believe that DL model assisted PR can be a practical and helpful tool for the BCDI community, being capable of simplifying the complex task of iterative phasing for hardly invertible datasets. Its use would help making the PR of highly strained crystal accessible to a larger number of BCDI users and expanding the applicability of the technique to many more interesting cases, previously inaccessible because of poor reconstructions.

\backmatter

\bmhead{Supplementary information} 
Supplementary information is available for this paper.
 
\section*{Declarations}

\begin{itemize}
\item Funding: \\
This project has been partly funded by the European Union’s Horizon 2020 Research and Innovation Programme under the Marie Sklodowska-Curie COFUND scheme with grant agreement No. 101034267 and the European Research Council (ERC) under the European's Horizon 2020 research and innovation programme (grant agreement No. 818823).
\item Availability of data and materials: The datasets used during the current study available from the corresponding author on reasonable request. Mentioned experimental datasets are provided with corresponding DOI. 
\item Code availability: The codes for this study are available and accessible via this link: 
\url{https://github.com/matteomasto/high_strain_CNN}
\end{itemize}



\bibliography{sn-bibliography}


\begin{thebibliography}{47}
\ifx \bisbn   \undefined \def \bisbn  #1{ISBN #1}\fi
\ifx \binits  \undefined \def \binits#1{#1}\fi
\ifx \bauthor  \undefined \def \bauthor#1{#1}\fi
\ifx \batitle  \undefined \def \batitle#1{#1}\fi
\ifx \bjtitle  \undefined \def \bjtitle#1{#1}\fi
\ifx \bvolume  \undefined \def \bvolume#1{\textbf{#1}}\fi
\ifx \byear  \undefined \def \byear#1{#1}\fi
\ifx \bissue  \undefined \def \bissue#1{#1}\fi
\ifx \bfpage  \undefined \def \bfpage#1{#1}\fi
\ifx \blpage  \undefined \def \blpage #1{#1}\fi
\ifx \burl  \undefined \def \burl#1{\textsf{#1}}\fi
\ifx \doiurl  \undefined \def \doiurl#1{\url{https://doi.org/#1}}\fi
\ifx \betal  \undefined \def \betal{\textit{et al.}}\fi
\ifx \binstitute  \undefined \def \binstitute#1{#1}\fi
\ifx \binstitutionaled  \undefined \def \binstitutionaled#1{#1}\fi
\ifx \bctitle  \undefined \def \bctitle#1{#1}\fi
\ifx \beditor  \undefined \def \beditor#1{#1}\fi
\ifx \bpublisher  \undefined \def \bpublisher#1{#1}\fi
\ifx \bbtitle  \undefined \def \bbtitle#1{#1}\fi
\ifx \bedition  \undefined \def \bedition#1{#1}\fi
\ifx \bseriesno  \undefined \def \bseriesno#1{#1}\fi
\ifx \blocation  \undefined \def \blocation#1{#1}\fi
\ifx \bsertitle  \undefined \def \bsertitle#1{#1}\fi
\ifx \bsnm \undefined \def \bsnm#1{#1}\fi
\ifx \bsuffix \undefined \def \bsuffix#1{#1}\fi
\ifx \bparticle \undefined \def \bparticle#1{#1}\fi
\ifx \barticle \undefined \def \barticle#1{#1}\fi
\bibcommenthead
\ifx \bconfdate \undefined \def \bconfdate #1{#1}\fi
\ifx \botherref \undefined \def \botherref #1{#1}\fi
\ifx \url \undefined \def \url#1{\textsf{#1}}\fi
\ifx \bchapter \undefined \def \bchapter#1{#1}\fi
\ifx \bbook \undefined \def \bbook#1{#1}\fi
\ifx \bcomment \undefined \def \bcomment#1{#1}\fi
\ifx \oauthor \undefined \def \oauthor#1{#1}\fi
\ifx \citeauthoryear \undefined \def \citeauthoryear#1{#1}\fi
\ifx \endbibitem  \undefined \def \endbibitem {}\fi
\ifx \bconflocation  \undefined \def \bconflocation#1{#1}\fi
\ifx \arxivurl  \undefined \def \arxivurl#1{\textsf{#1}}\fi
\csname PreBibitemsHook\endcsname

\bibitem[\protect\citeauthoryear{Miao et~al.}{2001}]{miao_approach_2001}
\begin{barticle}
\bauthor{\bsnm{Miao}, \binits{J.}},
\bauthor{\bsnm{Hodgson}, \binits{K.O.}},
\bauthor{\bsnm{Sayre}, \binits{D.}}:
\batitle{An approach to three-dimensional structures of biomolecules by using single-molecule diffraction images}.
\bjtitle{Proceedings of the National Academy of Sciences of the United States of America}
\bvolume{98}(\bissue{12}),
\bfpage{6641}--\blpage{6645}
(\byear{2001})
\doiurl{10.1073/pnas.111083998} .
Accessed 2020-04-01
\end{barticle}
\endbibitem

\bibitem[\protect\citeauthoryear{Robinson and Harder}{2009}]{Robinson2009}
\begin{barticle}
\bauthor{\bsnm{Robinson}, \binits{I.}},
\bauthor{\bsnm{Harder}, \binits{R.}}:
\batitle{Coherent x-ray diffraction imaging of strain at the nanoscale}.
\bjtitle{Nature Materials}
\bvolume{8}(\bissue{4}),
\bfpage{291}--\blpage{298}
(\byear{2009})
\doiurl{10.1038/nmat2400}
\end{barticle}
\endbibitem

\bibitem[\protect\citeauthoryear{Richard et~al.}{2022}]{Richard:te5091}
\begin{barticle}
\bauthor{\bsnm{Richard}, \binits{M.-I.}},
\bauthor{\bsnm{Labat}, \binits{S.}},
\bauthor{\bsnm{Dupraz}, \binits{M.}},
\bauthor{\bsnm{Li}, \binits{N.}},
\bauthor{\bsnm{Bellec}, \binits{E.}},
\bauthor{\bsnm{Boesecke}, \binits{P.}},
\bauthor{\bsnm{Djazouli}, \binits{H.}},
\bauthor{\bsnm{Eymery}, \binits{J.}},
\bauthor{\bsnm{Thomas}, \binits{O.}},
\bauthor{\bsnm{Sch{\"{u}}lli}, \binits{T.U.}},
\bauthor{\bsnm{Santala}, \binits{M.K.}},
\bauthor{\bsnm{Leake}, \binits{S.J.}}:
\batitle{{Bragg coherent diffraction imaging of single 20nm Pt particles at the ID01-EBS beamline of ESRF}}.
\bjtitle{Journal of Applied Crystallography}
\bvolume{55}(\bissue{3}),
\bfpage{621}--\blpage{625}
(\byear{2022})
\doiurl{10.1107/S1600576722002886}
\end{barticle}
\endbibitem

\bibitem[\protect\citeauthoryear{Carnis et~al.}{2021}]{Carnis2021}
\begin{barticle}
\bauthor{\bsnm{Carnis}, \binits{J.}},
\bauthor{\bsnm{Kshirsagar}, \binits{A.R.}},
\bauthor{\bsnm{Wu}, \binits{L.}},
\bauthor{\bsnm{Dupraz}, \binits{M.}},
\bauthor{\bsnm{Labat}, \binits{S.}},
\bauthor{\bsnm{Texier}, \binits{M.}},
\bauthor{\bsnm{Favre}, \binits{L.}},
\bauthor{\bsnm{Gao}, \binits{L.}},
\bauthor{\bsnm{Oropeza}, \binits{F.E.}},
\bauthor{\bsnm{Gazit}, \binits{N.}},
\bauthor{\bsnm{Almog}, \binits{E.}},
\bauthor{\bsnm{Campos}, \binits{A.}},
\bauthor{\bsnm{Micha}, \binits{J.-S.}},
\bauthor{\bsnm{Hensen}, \binits{E.J.M.}},
\bauthor{\bsnm{Leake}, \binits{S.J.}},
\bauthor{\bsnm{Sch{\"u}lli}, \binits{T.U.}},
\bauthor{\bsnm{Rabkin}, \binits{E.}},
\bauthor{\bsnm{Thomas}, \binits{O.}},
\bauthor{\bsnm{Poloni}, \binits{R.}},
\bauthor{\bsnm{Hofmann}, \binits{J.P.}},
\bauthor{\bsnm{Richard}, \binits{M.-I.}}:
\batitle{Twin boundary migration in an individual platinum nanocrystal during catalytic co oxidation}.
\bjtitle{Nature Communications}
\bvolume{12}(\bissue{1}),
\bfpage{5385}
(\byear{2021})
\doiurl{10.1038/s41467-021-25625-0}
\end{barticle}
\endbibitem

\bibitem[\protect\citeauthoryear{Hofmann et~al.}{2020}]{PhysRevMaterials.4.013801}
\begin{barticle}
\bauthor{\bsnm{Hofmann}, \binits{F.}},
\bauthor{\bsnm{Phillips}, \binits{N.W.}},
\bauthor{\bsnm{Das}, \binits{S.}},
\bauthor{\bsnm{Karamched}, \binits{P.}},
\bauthor{\bsnm{Hughes}, \binits{G.M.}},
\bauthor{\bsnm{Douglas}, \binits{J.O.}},
\bauthor{\bsnm{Cha}, \binits{W.}},
\bauthor{\bsnm{Liu}, \binits{W.}}:
\batitle{Nanoscale imaging of the full strain tensor of specific dislocations extracted from a bulk sample}.
\bjtitle{Phys. Rev. Mater.}
\bvolume{4},
\bfpage{013801}
(\byear{2020})
\doiurl{10.1103/PhysRevMaterials.4.013801}
\end{barticle}
\endbibitem

\bibitem[\protect\citeauthoryear{Godard}{2021}]{Godard2021}
\begin{barticle}
\bauthor{\bsnm{Godard}, \binits{P.}}:
\batitle{On the use of the scattering amplitude in coherent x-ray bragg diffraction imaging}.
\bjtitle{Journal of Applied Crystallography}
\bvolume{54},
\bfpage{797}--\blpage{802}
(\byear{2021})
\doiurl{10.1107/S1600576721003113}
\end{barticle}
\endbibitem

\bibitem[\protect\citeauthoryear{Yang et~al.}{2013}]{Yang2013}
\begin{botherref}
\oauthor{\bsnm{Yang}, \binits{W.}},
\oauthor{\bsnm{Huang}, \binits{X.}},
\oauthor{\bsnm{Harder}, \binits{R.}},
\oauthor{\bsnm{Clark}, \binits{J.N.}},
\oauthor{\bsnm{Robinson}, \binits{I.K.}},
\oauthor{\bsnm{Mao}, \binits{H.K.}}:
Coherent diffraction imaging of nanoscale strain evolution in a single crystal under high pressure.
Nature Communications
\textbf{4}
(2013)
\doiurl{10.1038/ncomms2661}
\end{botherref}
\endbibitem

\bibitem[\protect\citeauthoryear{Estandarte et~al.}{2018}]{IronGold2018}
\begin{botherref}
\oauthor{\bsnm{Estandarte}, \binits{A.K.C.}},
\oauthor{\bsnm{Lynch}, \binits{C.M.}},
\oauthor{\bsnm{Monteforte}, \binits{M.}},
\oauthor{\bsnm{Rawle}, \binits{J.}},
\oauthor{\bsnm{Nicklin}, \binits{C.}},
\oauthor{\bsnm{Robinson}, \binits{I.}}:
Bragg coherent diffraction imaging of iron diffusion into gold nanocrystals
(2018)
\doiurl{10.1088/1367-2630/aaebc1}
\end{botherref}
\endbibitem

\bibitem[\protect\citeauthoryear{Rochet et~al.}{2019}]{ROCHET2019169}
\begin{barticle}
\bauthor{\bsnm{Rochet}, \binits{A.}},
\bauthor{\bsnm{Suzana}, \binits{A.F.}},
\bauthor{\bsnm{Passos}, \binits{A.R.}},
\bauthor{\bsnm{Kalile}, \binits{T.}},
\bauthor{\bsnm{Berenguer}, \binits{F.}},
\bauthor{\bsnm{Santilli}, \binits{C.V.}},
\bauthor{\bsnm{Pulcinelli}, \binits{S.H.}},
\bauthor{\bsnm{Meneau}, \binits{F.}}:
\batitle{In situ reactor to image catalysts at work in three-dimensions by bragg coherent x-ray diffraction}.
\bjtitle{Catalysis Today}
\bvolume{336},
\bfpage{169}--\blpage{173}
(\byear{2019})
\doiurl{10.1016/j.cattod.2018.12.020} .
\bcomment{Selected papers from the 6th International Congress on Operando Spectroscopy}
\end{barticle}
\endbibitem

\bibitem[\protect\citeauthoryear{Dupraz et~al.}{2022}]{Dupraz2022}
\begin{barticle}
\bauthor{\bsnm{Dupraz}, \binits{M.}},
\bauthor{\bsnm{Li}, \binits{N.}},
\bauthor{\bsnm{Carnis}, \binits{J.}},
\bauthor{\bsnm{Wu}, \binits{L.}},
\bauthor{\bsnm{Labat}, \binits{S.}},
\bauthor{\bsnm{Chatelier}, \binits{C.}},
\bauthor{\bsnm{Poll}, \binits{R.}},
\bauthor{\bsnm{Hofmann}, \binits{J.P.}},
\bauthor{\bsnm{Almog}, \binits{E.}},
\bauthor{\bsnm{Leake}, \binits{S.J.}},
\bauthor{\bsnm{Watier}, \binits{Y.}},
\bauthor{\bsnm{Lazarev}, \binits{S.}},
\bauthor{\bsnm{Westermeier}, \binits{F.}},
\bauthor{\bsnm{Sprung}, \binits{M.}},
\bauthor{\bsnm{Hensen}, \binits{E.J.M.}},
\bauthor{\bsnm{Thomas}, \binits{O.}},
\bauthor{\bsnm{Rabkin}, \binits{E.}},
\bauthor{\bsnm{Richard}, \binits{M.-I.}}:
\batitle{Imaging the facet surface strain state of supported multi-faceted pt nanoparticles during reaction}.
\bjtitle{Nature Communications}
\bvolume{13}(\bissue{1}),
\bfpage{3003}
(\byear{2022})
\doiurl{10.1038/s41467-022-30592-1}
\end{barticle}
\endbibitem

\bibitem[\protect\citeauthoryear{Atlan et~al.}{2023}]{Atlan2023}
\begin{barticle}
\bauthor{\bsnm{Atlan}, \binits{C.}},
\bauthor{\bsnm{Chatelier}, \binits{C.}},
\bauthor{\bsnm{Martens}, \binits{I.}},
\bauthor{\bsnm{Dupraz}, \binits{M.}},
\bauthor{\bsnm{Viola}, \binits{A.}},
\bauthor{\bsnm{Li}, \binits{N.}},
\bauthor{\bsnm{Gao}, \binits{L.}},
\bauthor{\bsnm{Leake}, \binits{S.J.}},
\bauthor{\bsnm{Schülli}, \binits{T.U.}},
\bauthor{\bsnm{Eymery}, \binits{J.}},
\bauthor{\bsnm{Maillard}, \binits{F.}},
\bauthor{\bsnm{Richard}, \binits{M.-I.}}:
\batitle{Imaging the strain evolution of a platinum nanoparticle under electrochemical control}.
\bjtitle{Nature Materials}
\bvolume{22}(\bissue{6}),
\bfpage{754}--\blpage{761}
(\byear{2023})
\doiurl{10.1038/s41563-023-01528-x}
\end{barticle}
\endbibitem

\bibitem[\protect\citeauthoryear{Chatelier et~al.}{2024}]{Chatelier:2023}
\begin{barticle}
\bauthor{\bsnm{Chatelier}, \binits{C.}},
\bauthor{\bsnm{Atlan}, \binits{C.}},
\bauthor{\bsnm{Dupraz}, \binits{M.}},
\bauthor{\bsnm{Leake}, \binits{S.}},
\bauthor{\bsnm{Li}, \binits{N.}},
\bauthor{\bsnm{Schülli}, \binits{T.U.}},
\bauthor{\bsnm{Levi}, \binits{M.}},
\bauthor{\bsnm{Rabkin}, \binits{E.}},
\bauthor{\bsnm{Favre}, \binits{L.}},
\bauthor{\bsnm{Labat}, \binits{S.}},
\bauthor{\bsnm{Eymery}, \binits{J.}},
\bauthor{\bsnm{Richard}, \binits{M.I.}}:
\batitle{Unveiling core-shell structure formation in a ni3fe nanoparticle with in situ multi-bragg coherent diffraction imaging}.
\bjtitle{ACS Nano}
\bvolume{18},
\bfpage{13517}--\blpage{13527}
(\byear{2024})
\doiurl{10.1021/acsnano.3c11534}
\end{barticle}
\endbibitem

\bibitem[\protect\citeauthoryear{Gerchberg}{1972}]{Gerchberg1972APA}
\begin{barticle}
\bauthor{\bsnm{Gerchberg}, \binits{R.W.}}:
\batitle{A practical algorithm for the determination of phase from image and diffraction plane pictures}.
\bjtitle{Optik}
\bvolume{35},
\bfpage{237}--\blpage{246}
(\byear{1972})
\end{barticle}
\endbibitem

\bibitem[\protect\citeauthoryear{Fienup}{1978}]{Fienup:78}
\begin{barticle}
\bauthor{\bsnm{Fienup}, \binits{J.R.}}:
\batitle{Reconstruction of an object from the modulus of its fourier transform}.
\bjtitle{Opt. Lett.}
\bvolume{3}(\bissue{1}),
\bfpage{27}--\blpage{29}
(\byear{1978})
\doiurl{10.1364/OL.3.000027}
\end{barticle}
\endbibitem

\bibitem[\protect\citeauthoryear{Marchesini}{2007}]{Marchesini2007ARetrieval}
\begin{botherref}
\oauthor{\bsnm{Marchesini}, \binits{S.}}:
A unified evaluation of iterative projection algorithms for phase retrieval.
Review of Scientific Instruments
\textbf{78}(1)
(2007)
\doiurl{10.1063/1.2403783}
\end{botherref}
\endbibitem

\bibitem[\protect\citeauthoryear{Favre-Nicolin et~al.}{2020a}]{Favre-Nicolin2020PyNX:Operators}
\begin{barticle}
\bauthor{\bsnm{Favre-Nicolin}, \binits{V.}},
\bauthor{\bsnm{Girard}, \binits{G.}},
\bauthor{\bsnm{Leake}, \binits{S.}},
\bauthor{\bsnm{Carnis}, \binits{J.}},
\bauthor{\bsnm{Chushkin}, \binits{Y.}},
\bauthor{\bsnm{Kieffer}, \binits{J.}},
\bauthor{\bsnm{Paleo}, \binits{P.}},
\bauthor{\bsnm{Richard}, \binits{M.I.}}:
\batitle{{PyNX: High-performance computing toolkit for coherent X-ray imaging based on operators}}.
\bjtitle{Journal of Applied Crystallography}
\bvolume{53},
\bfpage{1404}--\blpage{1413}
(\byear{2020})
\doiurl{10.1107/S1600576720010985}
\end{barticle}
\endbibitem

\bibitem[\protect\citeauthoryear{Favre-Nicolin et~al.}{2020b}]{Favre-Nicolin2020}
\begin{botherref}
\oauthor{\bsnm{Favre-Nicolin}, \binits{V.}},
\oauthor{\bsnm{Leake}, \binits{S.}},
\oauthor{\bsnm{Chushkin}, \binits{Y.}}:
Free log-likelihood as an unbiased metric for coherent diffraction imaging.
Scientific Reports
\textbf{10}
(2020)
\doiurl{10.1038/s41598-020-57561-2}
\end{botherref}
\endbibitem

\bibitem[\protect\citeauthoryear{Zhao et~al.}{2023}]{Zhao}
\begin{barticle}
\bauthor{\bsnm{Zhao}, \binits{J.}},
\bauthor{\bsnm{Vartanyants}, \binits{I.A.}},
\bauthor{\bsnm{Zhang}, \binits{F.}}:
\batitle{Bragg coherent modulation imaging for highly strained nanocrystals: a numerical study}.
\bjtitle{Journal of Applied Crystallography}
\bvolume{56}(\bissue{5}),
\bfpage{1528}--\blpage{1536}
(\byear{2023})
\doiurl{10.1107/S1600576723007720}
{\href{https://arxiv.org/abs/https://onlinelibrary.wiley.com/doi/pdf/10.1107/S1600576723007720}{{https://onlinelibrary.wiley.com/doi/pdf/10.1107/S1600576723007720}}}
\end{barticle}
\endbibitem

\bibitem[\protect\citeauthoryear{}{2025}]{Zhao_unpublished}
\begin{botherref}
Bragg coherent modulation imaging of highly-strained nanocrystals (submitted)
(2025)
\end{botherref}
\endbibitem

\bibitem[\protect\citeauthoryear{Wang et~al.}{2020}]{Wang_2020}
\begin{barticle}
\bauthor{\bsnm{Wang}, \binits{Z.}},
\bauthor{\bsnm{Gorobtsov}, \binits{O.}},
\bauthor{\bsnm{Singer}, \binits{A.}}:
\batitle{An algorithm for bragg coherent x-ray diffractive imaging of highly strained nanocrystals}.
\bjtitle{New Journal of Physics}
\bvolume{22}(\bissue{1}),
\bfpage{013021}
(\byear{2020})
\doiurl{10.1088/1367-2630/ab61db}
\end{barticle}
\endbibitem

\bibitem[\protect\citeauthoryear{Newton et~al.}{2010}]{NewtonStrain}
\begin{barticle}
\bauthor{\bsnm{Newton}, \binits{M.C.}},
\bauthor{\bsnm{Harder}, \binits{R.}},
\bauthor{\bsnm{Huang}, \binits{X.}},
\bauthor{\bsnm{Xiong}, \binits{G.}},
\bauthor{\bsnm{Robinson}, \binits{I.K.}}:
\batitle{Phase retrieval of diffraction from highly strained crystals}.
\bjtitle{Phys. Rev. B}
\bvolume{82},
\bfpage{165436}
(\byear{2010})
\doiurl{10.1103/PhysRevB.82.165436}
\end{barticle}
\endbibitem

\bibitem[\protect\citeauthoryear{Minkevich et~al.}{2008}]{Minkevich2008}
\begin{botherref}
\oauthor{\bsnm{Minkevich}, \binits{A.A.}},
\oauthor{\bsnm{Baumbach}, \binits{T.}},
\oauthor{\bsnm{Gailhanou}, \binits{M.}},
\oauthor{\bsnm{Thomas}, \binits{O.}}:
Applicability of an iterative inversion algorithm to the diffraction patterns from inhomogeneously strained crystals.
Physical Review B - Condensed Matter and Materials Physics
\textbf{78}
(2008)
\doiurl{10.1103/PhysRevB.78.174110}
\end{botherref}
\endbibitem

\bibitem[\protect\citeauthoryear{Cherukara et~al.}{2018}]{Cherukara2018Real-timeNetworks}
\begin{botherref}
\oauthor{\bsnm{Cherukara}, \binits{M.J.}},
\oauthor{\bsnm{Nashed}, \binits{Y.S.G.}},
\oauthor{\bsnm{Harder}, \binits{R.J.}}:
{Real-time coherent diffraction inversion using deep generative networks}.
Scientific Reports
\textbf{8}(1)
(2018)
\doiurl{10.1038/s41598-018-34525-1}
\end{botherref}
\endbibitem

\bibitem[\protect\citeauthoryear{Scheinker and Pokharel}{2020}]{Scheinker2020AdaptiveImaging}
\begin{botherref}
\oauthor{\bsnm{Scheinker}, \binits{A.}},
\oauthor{\bsnm{Pokharel}, \binits{R.}}:
{Adaptive 3D convolutional neural network-based reconstruction method for 3D coherent diffraction imaging}.
Journal of Applied Physics
\textbf{128}(18)
(2020)
\doiurl{10.1063/5.0014725}
\end{botherref}
\endbibitem

\bibitem[\protect\citeauthoryear{Wu et~al.}{2021a}]{Wu:cw5029}
\begin{barticle}
\bauthor{\bsnm{Wu}, \binits{L.}},
\bauthor{\bsnm{Juhas}, \binits{P.}},
\bauthor{\bsnm{Yoo}, \binits{S.}},
\bauthor{\bsnm{Robinson}, \binits{I.}}:
\batitle{{Complex imaging of phase domains by deep neural networks}}.
\bjtitle{IUCrJ}
\bvolume{8}(\bissue{1}),
\bfpage{12}--\blpage{21}
(\byear{2021})
\doiurl{10.1107/S2052252520013780}
\end{barticle}
\endbibitem

\bibitem[\protect\citeauthoryear{Wu et~al.}{2021b}]{Wu2021Three-dimensionalNetworks}
\begin{botherref}
\oauthor{\bsnm{Wu}, \binits{L.}},
\oauthor{\bsnm{Yoo}, \binits{S.}},
\oauthor{\bsnm{Suzana}, \binits{A.F.}},
\oauthor{\bsnm{Assefa}, \binits{T.A.}},
\oauthor{\bsnm{Diao}, \binits{J.}},
\oauthor{\bsnm{Harder}, \binits{R.J.}},
\oauthor{\bsnm{Cha}, \binits{W.}},
\oauthor{\bsnm{Robinson}, \binits{I.K.}}:
{Three-dimensional coherent X-ray diffraction imaging via deep convolutional neural networks}.
npj Computational Materials
\textbf{7}(1)
(2021)
\doiurl{10.1038/s41524-021-00644-z}
\end{botherref}
\endbibitem

\bibitem[\protect\citeauthoryear{Yao et~al.}{2022}]{Yao2022AutoPhaseNN:Imaging}
\begin{botherref}
\oauthor{\bsnm{Yao}, \binits{Y.}},
\oauthor{\bsnm{Chan}, \binits{H.}},
\oauthor{\bsnm{Sankaranarayanan}, \binits{S.}},
\oauthor{\bsnm{Balaprakash}, \binits{P.}},
\oauthor{\bsnm{Harder}, \binits{R.J.}},
\oauthor{\bsnm{Cherukara}, \binits{M.J.}}:
{AutoPhaseNN: unsupervised physics-aware deep learning of 3D nanoscale Bragg coherent diffraction imaging}.
npj Computational Materials
\textbf{8}(1)
(2022)
\doiurl{10.1038/s41524-022-00803-w}
\end{botherref}
\endbibitem

\bibitem[\protect\citeauthoryear{Yu et~al.}{2024}]{Yu2024}
\begin{botherref}
\oauthor{\bsnm{Yu}, \binits{X.}},
\oauthor{\bsnm{Wu}, \binits{L.}},
\oauthor{\bsnm{Lin}, \binits{Y.}},
\oauthor{\bsnm{Diao}, \binits{J.}},
\oauthor{\bsnm{Liu}, \binits{J.}},
\oauthor{\bsnm{Hallmann}, \binits{J.}},
\oauthor{\bsnm{Boesenberg}, \binits{U.}},
\oauthor{\bsnm{Lu}, \binits{W.}},
\oauthor{\bsnm{Möller}, \binits{J.}},
\oauthor{\bsnm{Scholz}, \binits{M.}},
\oauthor{\bsnm{Zozulya}, \binits{A.}},
\oauthor{\bsnm{Madsen}, \binits{A.}},
\oauthor{\bsnm{Assefa}, \binits{T.}},
\oauthor{\bsnm{Bozin}, \binits{E.S.}},
\oauthor{\bsnm{Cao}, \binits{Y.}},
\oauthor{\bsnm{You}, \binits{H.}},
\oauthor{\bsnm{Sheyfer}, \binits{D.}},
\oauthor{\bsnm{Rosenkranz}, \binits{S.}},
\oauthor{\bsnm{Marks}, \binits{S.D.}},
\oauthor{\bsnm{Evans}, \binits{P.G.}},
\oauthor{\bsnm{Keen}, \binits{D.A.}},
\oauthor{\bsnm{He}, \binits{X.}},
\oauthor{\bsnm{Božović}, \binits{I.}},
\oauthor{\bsnm{Dean}, \binits{M.P.M.}},
\oauthor{\bsnm{Yoo}, \binits{S.}},
\oauthor{\bsnm{Robinson}, \binits{I.K.}}:
Ultrafast bragg coherent diffraction imaging of epitaxial thin films using deep complex-valued neural networks.
npj Computational Materials
\textbf{10}
(2024)
\doiurl{10.1038/s41524-024-01208-7}
\end{botherref}
\endbibitem

\bibitem[\protect\citeauthoryear{Maddali et~al.}{2023}]{Maddali2023}
\begin{botherref}
\oauthor{\bsnm{Maddali}, \binits{S.}},
\oauthor{\bsnm{Frazer}, \binits{T.D.}},
\oauthor{\bsnm{Delegan}, \binits{N.}},
\oauthor{\bsnm{Harmon}, \binits{K.J.}},
\oauthor{\bsnm{Sullivan}, \binits{S.E.}},
\oauthor{\bsnm{Allain}, \binits{M.}},
\oauthor{\bsnm{Cha}, \binits{W.}},
\oauthor{\bsnm{Dibos}, \binits{A.}},
\oauthor{\bsnm{Poudyal}, \binits{I.}},
\oauthor{\bsnm{Kandel}, \binits{S.}},
\oauthor{\bsnm{Nashed}, \binits{Y.S.G.}},
\oauthor{\bsnm{Heremans}, \binits{F.J.}},
\oauthor{\bsnm{You}, \binits{H.}},
\oauthor{\bsnm{Cao}, \binits{Y.}},
\oauthor{\bsnm{Hruszkewycz}, \binits{S.O.}}:
Concurrent multi-peak bragg coherent x-ray diffraction imaging of 3d nanocrystal lattice displacement via global optimization.
npj Computational Materials
\textbf{9}
(2023)
\doiurl{10.1038/s41524-023-01022-7}
\end{botherref}
\endbibitem

\bibitem[\protect\citeauthoryear{Özgün Çiçek et~al.}{2016}]{3DUnet2016}
\begin{botherref}
\oauthor{\bsnm{Çiçek}},
\oauthor{\bsnm{Abdulkadir}, \binits{A.}},
\oauthor{\bsnm{Lienkamp}, \binits{S.S.}},
\oauthor{\bsnm{Brox}, \binits{T.}},
\oauthor{\bsnm{Ronneberger}, \binits{O.}}:
3D U-Net: Learning Dense Volumetric Segmentation from Sparse Annotation
(2016).
\url{https://arxiv.org/abs/1606.06650}
\end{botherref}
\endbibitem

\bibitem[\protect\citeauthoryear{Mahmud et~al.}{2023}]{3DUnetSeg2023}
\begin{botherref}
\oauthor{\bsnm{Mahmud}, \binits{B.U.}},
\oauthor{\bsnm{Hong}, \binits{G.Y.}},
\oauthor{\bsnm{Mamun}, \binits{A.A.}},
\oauthor{\bsnm{Ping}, \binits{E.P.}},
\oauthor{\bsnm{Wu}, \binits{Q.}}:
Deep learning-based segmentation of 3d volumetric image and microstructural analysis
(2023)
\doiurl{10.3390/s23052640}
\end{botherref}
\endbibitem

\bibitem[\protect\citeauthoryear{Xu et~al.}{2024}]{Xu2024}
\begin{botherref}
\oauthor{\bsnm{Xu}, \binits{G.}},
\oauthor{\bsnm{Wang}, \binits{X.}},
\oauthor{\bsnm{Wu}, \binits{X.}},
\oauthor{\bsnm{Leng}, \binits{X.}},
\oauthor{\bsnm{Xu}, \binits{Y.}}:
Development of skip connection in deep neural networks for computer vision and medical image analysis: A survey
(2024)
\end{botherref}
\endbibitem

\bibitem[\protect\citeauthoryear{Lim et~al.}{2021}]{Lim2021ADiffraction}
\begin{botherref}
\oauthor{\bsnm{Lim}, \binits{B.}},
\oauthor{\bsnm{Bellec}, \binits{E.}},
\oauthor{\bsnm{Dupraz}, \binits{M.}},
\oauthor{\bsnm{Leake}, \binits{S.}},
\oauthor{\bsnm{Resta}, \binits{A.}},
\oauthor{\bsnm{Coati}, \binits{A.}},
\oauthor{\bsnm{Sprung}, \binits{M.}},
\oauthor{\bsnm{Almog}, \binits{E.}},
\oauthor{\bsnm{Rabkin}, \binits{E.}},
\oauthor{\bsnm{Schulli}, \binits{T.}},
\oauthor{\bsnm{Richard}, \binits{M.I.}}:
{A convolutional neural network for defect classification in Bragg coherent X-ray diffraction}.
npj Computational Materials
\textbf{7}(1)
(2021)
\doiurl{10.1038/s41524-021-00583-9}
\end{botherref}
\endbibitem

\bibitem[\protect\citeauthoryear{Favre-Nicolin et~al.}{2011}]{Favre_scattering}
\begin{barticle}
\bauthor{\bsnm{Favre-Nicolin}, \binits{V.}},
\bauthor{\bsnm{Coraux}, \binits{J.}},
\bauthor{\bsnm{Richard}, \binits{M.-I.}},
\bauthor{\bsnm{Renevier}, \binits{H.}}:
\batitle{{Fast computation of scattering maps of nanostructures using graphical processing units}}.
\bjtitle{Journal of Applied Crystallography}
\bvolume{44}(\bissue{3}),
\bfpage{635}--\blpage{640}
(\byear{2011})
\doiurl{10.1107/S0021889811009009}
\end{barticle}
\endbibitem

\bibitem[\protect\citeauthoryear{Masto et~al.}{2024}]{Masto:yr5131}
\begin{barticle}
\bauthor{\bsnm{Masto}, \binits{M.}},
\bauthor{\bsnm{Favre-Nicolin}, \binits{V.}},
\bauthor{\bsnm{Leake}, \binits{S.}},
\bauthor{\bsnm{Sch{\"{u}}lli}, \binits{T.}},
\bauthor{\bsnm{Richard}, \binits{M.-I.}},
\bauthor{\bsnm{Bellec}, \binits{E.}}:
\batitle{{Patching-based deep-learning model for the inpainting of Bragg coherent diffraction patterns affected by detector gaps}}.
\bjtitle{Journal of Applied Crystallography}
\bvolume{57}(\bissue{4}),
\bfpage{966}--\blpage{974}
(\byear{2024})
\doiurl{10.1107/S1600576724004163}
\end{barticle}
\endbibitem

\bibitem[\protect\citeauthoryear{Özgün Çiçek et~al.}{2016}]{3DUnet}
\begin{botherref}
\oauthor{\bsnm{Çiçek}},
\oauthor{\bsnm{Abdulkadir}, \binits{A.}},
\oauthor{\bsnm{Lienkamp}, \binits{S.S.}},
\oauthor{\bsnm{Brox}, \binits{T.}},
\oauthor{\bsnm{Ronneberger}, \binits{O.}}:
3D U-Net: Learning Dense Volumetric Segmentation from Sparse Annotation
(2016).
\url{https://arxiv.org/abs/1606.06650}
\end{botherref}
\endbibitem

\bibitem[\protect\citeauthoryear{Chen et~al.}{2017}]{Chen2017RethinkingSegmentation}
\begin{botherref}
\oauthor{\bsnm{Chen}, \binits{L.-C.}},
\oauthor{\bsnm{Papandreou}, \binits{G.}},
\oauthor{\bsnm{Schroff}, \binits{F.}},
\oauthor{\bsnm{Adam}, \binits{H.}}:
{Rethinking Atrous Convolution for Semantic Image Segmentation}
(2017)
\end{botherref}
\endbibitem

\bibitem[\protect\citeauthoryear{Abadi}{2016}]{abadi2016tensorflow}
\begin{bchapter}
\bauthor{\bsnm{Abadi}, \binits{M.}}:
\bctitle{Tensorflow: learning functions at scale}.
In: \bbtitle{Proceedings of the 21st ACM SIGPLAN International Conference on Functional Programming},
pp. \bfpage{1}--\blpage{1}
(\byear{2016})
\end{bchapter}
\endbibitem

\bibitem[\protect\citeauthoryear{Kingma and Ba}{2017}]{kingma2017adam}
\begin{botherref}
\oauthor{\bsnm{Kingma}, \binits{D.P.}},
\oauthor{\bsnm{Ba}, \binits{J.}}:
Adam: A Method for Stochastic Optimization
(2017)
\end{botherref}
\endbibitem

\bibitem[\protect\citeauthoryear{Guizar-Sicairos and Fienup}{2012}]{Guizar-Sicairos:12}
\begin{barticle}
\bauthor{\bsnm{Guizar-Sicairos}, \binits{M.}},
\bauthor{\bsnm{Fienup}, \binits{J.R.}}:
\batitle{Understanding the twin-image problem in phase retrieval}.
\bjtitle{J. Opt. Soc. Am. A}
\bvolume{29}(\bissue{11}),
\bfpage{2367}--\blpage{2375}
(\byear{2012})
\doiurl{10.1364/JOSAA.29.002367}
\end{barticle}
\endbibitem

\bibitem[\protect\citeauthoryear{Zhang et~al.}{2024}]{Sun_symmetry2024}
\begin{botherref}
\oauthor{\bsnm{Zhang}, \binits{W.}},
\oauthor{\bsnm{Wan}, \binits{Y.}},
\oauthor{\bsnm{Zhuang}, \binits{Z.}},
\oauthor{\bsnm{Sun}, \binits{J.}}:
What is Wrong with End-to-End Learning for Phase Retrieval?
(2024).
\url{https://arxiv.org/abs/2403.15448}
\end{botherref}
\endbibitem

\bibitem[\protect\citeauthoryear{Richard et~al.}{2025a}]{data1}
\begin{botherref}
\oauthor{\bsnm{Richard}, \binits{M.-I.}},
\oauthor{\bsnm{Bellec}, \binits{E.}},
\oauthor{\bsnm{Chatelier}, \binits{C.}},
\oauthor{\bsnm{Atlan}, \binits{C.}},
\oauthor{\bsnm{Zhao}, \binits{J.}},
\oauthor{\bsnm{Viola}, \binits{A.}},
\oauthor{\bsnm{Grimes}, \binits{M.}},
\oauthor{\bsnm{Olson}, \binits{K.}}:
Bragg coherent diffraction imaging of Pt nanoparticles on sapphire (Version 1) [Dataset].
European Synchrotron Radiation Facility.
Includes data from 0001\_scan\_65
(2025).
\doiurl{10.15151/ESRF-DC-2184066699} .
\url{https://doi.org/10.15151/ESRF-DC-2184066699}
\end{botherref}
\endbibitem

\bibitem[\protect\citeauthoryear{Richard et~al.}{2025b}]{data2}
\begin{botherref}
\oauthor{\bsnm{Richard}, \binits{M.-I.}},
\oauthor{\bsnm{Bouita}, \binits{M.}},
\oauthor{\bsnm{Leake}, \binits{S.}},
\oauthor{\bsnm{Atlan}, \binits{C.}},
\oauthor{\bsnm{Bellec}, \binits{E.}},
\oauthor{\bsnm{Olson}, \binits{K.}},
\oauthor{\bsnm{Khater}, \binits{P.}}:
Bragg coherent diffraction imaging of a PdPt nanoparticle (Version 1) [Dataset].
European Synchrotron Radiation Facility.
Includes data from scan\_4
(2025).
\doiurl{10.15151/ESRF-DC-2014789918} .
\url{https://doi.org/10.15151/ESRF-DC-2014789918}
\end{botherref}
\endbibitem

\bibitem[\protect\citeauthoryear{Marchesini et~al.}{2003}]{Marchesini2003}
\begin{botherref}
\oauthor{\bsnm{Marchesini}, \binits{S.}},
\oauthor{\bsnm{He}, \binits{H.}},
\oauthor{\bsnm{Chapman}, \binits{N.}},
\oauthor{\bsnm{Hau-Riege}, \binits{P.}},
\oauthor{\bsnm{Noy}, \binits{A.}},
\oauthor{\bsnm{Howells}, \binits{R.}},
\oauthor{\bsnm{Weierstall}, \binits{U.}},
\oauthor{\bsnm{Spence}, \binits{H.}}:
X-ray image reconstruction from a diffraction pattern alone.
Physical Review B - Condensed Matter and Materials Physics
\textbf{68}
(2003)
\doiurl{10.1103/PhysRevB.68.140101}
\end{botherref}
\endbibitem

\bibitem[\protect\citeauthoryear{Crimmins et~al.}{1990}]{Crimmins:90}
\begin{barticle}
\bauthor{\bsnm{Crimmins}, \binits{T.R.}},
\bauthor{\bsnm{Fienup}, \binits{J.R.}},
\bauthor{\bsnm{Thelen}, \binits{B.J.}}:
\batitle{Improved bounds on object support from autocorrelation support and application to phase retrieval}.
\bjtitle{J. Opt. Soc. Am. A}
\bvolume{7}(\bissue{1}),
\bfpage{3}--\blpage{13}
(\byear{1990})
\doiurl{10.1364/JOSAA.7.000003}
\end{barticle}
\endbibitem

\bibitem[\protect\citeauthoryear{Atlan}{2023}]{atlan2023cdiutils}
\begin{botherref}
\oauthor{\bsnm{Atlan}, \binits{C.}}:
Cdiutils: A Python Package for X-ray Bragg Coherent Diffraction Imaging Processing, Analysis and Visualisation Workflows.
\doiurl{10.5281/zenodo.7656853} .
\url{https://github.com/clatlan/cdiutils}
\end{botherref}
\endbibitem

\bibitem[\protect\citeauthoryear{Favre-Nicolin et~al.}{2020}]{FreeLLK}
\begin{botherref}
\oauthor{\bsnm{Favre-Nicolin}, \binits{V.}},
\oauthor{\bsnm{Leake}, \binits{S.}},
\oauthor{\bsnm{Chushkin}, \binits{Y.}}:
Free log-likelihood as an unbiased metric for coherent diffraction imaging.
Scientific Reports
\textbf{10}
(2020)
\doiurl{10.1038/s41598-020-57561-2}
\end{botherref}
\endbibitem

\end{thebibliography}

\end{document}